\RequirePackage{scrlfile}
\PreventPackageFromLoading[]{natbib}

\documentclass[nonatbib,5p]{elsarticle}
\makeatletter

\let\c@author\relax
\makeatother

\usepackage{dblfloatfix}    
\usepackage{etoolbox,siunitx}
\robustify\bfseries

\usepackage{booktabs}
\usepackage{pdfpages} 

\usepackage{url}
\usepackage{caption}
\usepackage{subcaption}
\usepackage[inline]{enumitem}
\newlist{mylist}{enumerate*}{1}
\setlist[mylist]{label=(\roman*)}
\usepackage{color}
\usepackage{graphicx}
\usepackage{algorithm}
\usepackage[noend]{algpseudocode}
\usepackage{multirow}
\usepackage{textcomp} 
\usepackage[none]{hyphenat} 

\usepackage[
singlelinecheck=false 
]{caption}

\usepackage{amsfonts}       
\usepackage{amsmath, bm}    
\usepackage[utf8]{inputenc}
\usepackage[english]{babel}

\RequirePackage[					
	backend=biber,		
	bibencoding=utf8,				
	style=authoryear,		
	natbib=false,					
	dashed=false,                   
	hyperref=true,					
	backref=false,					
	isbn=false,						
	url=false,						
	doi=true,						
	urldate=long,					
	uniquename=false,               
	uniquelist=minyear,
	maxbibnames=5,%
	minbibnames=1,%
	maxcitenames=2,%
	mincitenames=1%
]{biblatex}
\definecolor{mygreen}{rgb}{0.06, 0.48, 0.023}
\definecolor{myblue}{rgb}{0.13, 0.59, 0.82}

\usepackage{hyperref} 
\hypersetup{					
	plainpages=false,			
	colorlinks=true,			
	pdfborder={0 0 0},			
	breaklinks=true,			
	bookmarksnumbered=true,		%
	bookmarksopen=true			%
}
    
\hypersetup{
    linkcolor = black,
	citecolor = myblue,
	urlcolor = blue,
}

\usepackage{fancyhdr}
\begin{document}
\sloppy 
\renewcommand*{\today}{March 9, 2019}

\hypersetup{allcolors = myblue}

\title{Detecting and monitoring long-term landslides in urbanized areas with nighttime light data and multi-seasonal Landsat imagery across Taiwan from 1998 to 2017}

\journal{Remote Sensing of Environment (accepted)}
\author[add1,add2,add3]{Tzu-Hsin Karen Chen\corref{corr}}
\cortext[corr]{Corresponding author}
\ead{thc@envs.au.dk}

\author[add3]{Alexander V. Prishchepov}
\author[add3]{Rasmus Fensholt}
\author[add1,add2]{Clive E. Sabel}

\address[add1]{Department of Environmental Science, Aarhus University, Frederiksborgvej 399, DK-4000 Roskilde, Denmark}
\address[add2]{Danish Big Data Centre for Environment and Health (BERTHA), Aarhus University, DK-4000 Roskilde, Denmark}
\address[add3]{Department of Geosciences and Natural Resource Management (IGN), University of Copenhagen, DK-1350 København K, Denmark}

\begin{abstract}

Monitoring long-term landslide activity is of importance for risk assessment and land management. Daytime airborne drones or very high-resolution optical satellites are often used to create landslide maps. However, such imagery comes at a high cost, making long-term risk analysis cost-prohibitive. Despite the widespread use of open-access 30m Landsat imagery, their utility for landslide detection is often limited due to low classification accuracy. One of the major challenges is to separate landslides from other anthropogenic disturbances. Here, we produce landslide maps retrospectively from 1998 to 2017 for landslide-prone and highly populated Taiwan (35,874 km\textsuperscript{2}). To improve classification accuracy of landslides, we integrate nighttime light imagery from the Defense Meteorological Satellite Program (DMSP) and the Visible Infrared Imaging Radiometer Suite (VIIRS), with multi-seasonal daytime optical Landsat time-series, and digital elevation data from the Advanced Spaceborne Thermal Emission and Reflection Radiometer (ASTER). We employed a non-parametric machine-learning classifier, random forest, to classify the satellite imagery. The classifier was trained with data from three years (2005, 2010, and 2015), and was validated with an independent reference sample from twelve years. Our results demonstrated that combining nighttime light data and multi-seasonal imagery significantly improved the classification (p$<$0.001), compared to conventional methods based on single-season optical imagery. The results confirmed that the developed classification model enabled mapping of landslides across Taiwan over a long period with annual overall accuracy varying between 96\% and 97\%, user's and producer's accuracies between 73\% and 86\%. Spatiotemporal analysis of the landslide inventories from 1998 to 2017 revealed different temporal patterns of landslide activities, showing those areas where landslides were persistent and other areas where landslides tended to reoccur after vegetation regrowth. In sum, we provide a robust method to detect long-term landslide activities based on freely available satellite imagery, which can be applied elsewhere. Our mapping effort of landslide spatiotemporal patterns is expected to be of high importance in developing effective landslide remediation strategies. Please find the published version at: \url{https://doi.org/10.1016/j.rse.2019.03.013}

\end{abstract}


\maketitle              

\section{Introduction}

From 1990 to 2017, landslides resulted in many casualties and caused roughly USD 4.5 billion of economic losses worldwide (\cite{EMDAT2018}). Approximately 58\% of deaths and 69\% of economic losses occurred in East and Southeast Asia (\cite{EMDAT2018}). East and Southeast Asia, one of the fastest urbanizing regions in the world, is currently experiencing rapid built-up area expansion (\cite{Schneider2015}). Recent deadly events causing many thousands of casualties in densely populated regions, include the 2013 Uttarakhand landslides in India (183 persons/km\textsuperscript{2}) (\cite{Martha2015}), the 2008 Sichuan landslides in China (170 persons/km\textsuperscript{2}) (\cite{Chigira2010}) and the 2009 Kaohsiung landslides in Taiwan (445 persons/km\textsuperscript{2}) (\cite{Tsou2011}). 

While in those regions the influence of population growth on landslide occurrence is inevitable (\cite{Petley2010}), historical landslide inventories are of great importance for supporting mitigation and adaptation strategies, such as monitoring landslide susceptible areas (\cite{Althuwaynee2012}). Historical landslide maps that cover longer periods can characterize crucial spatiotemporal patterns of landslide activities. Cumulative occurrence of landslides, for instance, is considered as a predictor of landslides (\cite{Chuang2018}). Other temporal characteristics, such as persistency, and reoccurrence rate after revegetation of former landslide sites, also reflect multiple factors such as the type and size of the landslide, the depths of the sliding plane, and the hydrogeological characteristics (\cite{Behling2014}). Thus, the knowledge gained from long-term landslide maps is useful for developing pathways for sustainable land use in landslide-prone areas (\cite{Fell2008}).

However, historical landslide maps are largely unavailable because of mapping challenges. Mapping of landslide occurrence traditionally relies on field surveys and visual interpretation of aerial photos. Semi-automatic pixel-based and object-based image classification methods have also been developed to extract information about landslides from very high-resolution (VHR) satellite imagery (e.g., Quickbird, IKONOS, WorldView) (\cite{Hervas2003,Pradhan2016, Stumpf2011,Keyport2018}). Studies have shown that object-based classification of images at $<$1m resolution, such as Quickbird and GeoEye-1, which utilize the geometry and texture of objects, can yield high overall accuracy (above 85\%) in landslide mapping (\cite{Li2015,Stumpf2011}). Nevertheless, retrospective mapping of landslide occurrence from VHR imagery is costly and does not allow tracing landslides before the late 1990s, when the first commercial VHR imagery became accessible. Repetitive observations with dense satellite time-series such as from Aqua/Terra MODIS at 250m resolution are cost-free but impose substantial limitations for detecting landslide due to the low spatial resolution.

In contrast, freely accessible medium-resolution satellite imagery, such as Landsat (30m) and SPOT (approximately 10m) imagery provide temporal coverage over several decades and are widely used to study land-cover change. Normalized Difference Vegetation Index (NDVI) and Normalized Difference Water Index (NDWI) derived from multispectral imagery are commonly used to differentiate rock-, soil-, and mud-covered landslides from vegetation and water bodies (\cite{Chen2013,Singh2016,Chen2018ppgis}). However, Landsat-like imagery alone usually resulted in moderate classification accuracies (\cite{Barlow2003,Yu2017}). Even when geomorphological parameters derived from a digital elevation model (DEM) were used as ancillary data, the commission error and omission error for landslide detection can be as high as 30\% (\cite{Hong2016,Yu2017}). The major challenge related to the use of Landsat-like multispectral imagery comes from the separation between landslides and the surrounding environment, as spectral signatures from landslide occurrences and anthropogenic landscapes are often similar. Human activities such as agricultural land use, forest clearcutting and constructions in hazard-prone zones in East and Southeast Asia, therefore, pose specific challenges regarding the accuracy of landslide detection (\cite{Borghuis2007,Martha2012}). 

Studies show that the applications of multi-temporal imagery can enhance land cover classification, such as fusing multi-date imagery, extracting time-series trajectories, and incorporating multiple seasonal features (\cite{Oetter2001,Guerschman2003,Khatami2016,Kennedy2010}). Fusing multi-date images can substitute missing information (e.g. clouds in optical satellite imagery) (\cite{Pohl1998}). While many landslide mapping efforts were made on event-based bi-temporal change detection (\cite{Hoelbling2015,Tsai2010,Rau2007}), methods which use time-series imagery for monitoring ongoing processes have been developed in the past decade. For instance, NDVI time-series were used for adaptive change calculation, allowing the separation between permanently non-vegetated and post-event landslide areas in different geographic settings (\cite{Behling2014,Golovko2017}). To monitor the geomorphological process of slow-movement landslides, time-series based optical imagery approaches have also been developed for retrieving surface displacements (\cite{Pham2018,Stumpf2017}). Synthetic Aperture Radar interferometry (InSAR), known for its ability in dealing with atmospheric noise and monitoring deformation, was also found to perform better using time-series analysis of landslides (\cite{Hooper2012,Dong2018}). However, time-series approaches require a high imagery frequency (\cite{Behling2016,Kennedy2010}), imposing difficulties for long-term annual landslide mapping. While multi-seasonal Landsat images have been used for long-term land-cover mapping and change detection (\cite{Gradinaru2017,Makhamreh2018,Prishchepov2012}), their utility for constructing historical landslide maps are underexplored. We hypothesize that multi-seasonal imagery should improve classification accuracy, because landslides may occur or expand during the rainy season and expose bare soil while agricultural fields remains green. 

We also hypothesize that ancillary nighttime light imagery could be applied to assess the spatial distribution of landslides. While optical imagery distinguishes landslide from vegetation, nighttime light data could separate landslide from anthropogenic activities interlinked with light use. Nighttime light data are tightly associated with population density, income level and economic activities (\cite{Doll2006,Jean2016}). For instance, MODIS imagery combined with nighttime light data enhanced classification accuracy of urban areas (\cite{Sharma2016}). Thus, we assume that by adding nighttime light data a classifier can better exclude such human activities from the areas affected by landslides, thereby overcoming the limitations of optical daytime reflectance data. The Defense Meteorological Satellite Program (DMSP) nighttime light data in combination with data from the more recent Visible Infrared Imaging Radiometer Suite (VIIRS) are available retrospectively, which combined with Landsat imagery provide a great potential to map landslides into the past. We are unaware of any studies that tested the suitability of nighttime light data to support automated landslide detection.

\begin{table*}
\caption{Remote sensing variables used in this study and the spatial resolution of each dataset.}
\centering
\setlength{\tabcolsep}{5pt}
\centering
\label{tab:1}
\scriptsize
\begin{tabular}{lll}
\hline
Metric                                & Description                                                                                                                               & Spatial   resolution \\ \hline
Seasonal features                     & Seasonal mean values of winter, spring, and   summer of bands (Blue, green, red, NIR, SWIR1, SWIR2), & 30m                  \\
                                      & (\cite{Nichol2005}), NDVI (\cite{Martha2011}), and NDWI (\cite{Wu2013}), where                                               &                      \\
                                      & NDVI=$\frac{NIR-red}{NIR+red}$                                                                                                            &                      \\
                                      & NDWI=$\frac{green-SWIR1}{green+SWIR1}$                                                                                                    &                      \\
Slope                                 & Degree values between 0 and 90 derived from DEM (\cite{Borghuis2007}).                                                   & 30m                  \\
\multirow{2}{*}{Nighttime light data} & \multirow{2}{*}{Values rescaled to 0 (no light) $–$ 63 (top   luminosity)}                                                                & about   900m (DMSP)  \\
                                      &                                                                                                                                           & about   450m (VIIRS) \\ \hline
\end{tabular}
\end{table*}

In this context, our major objective was to develop an automated approach for retrospective annual landslide mapping based on open-access satellite data (Landsat, DMSP and VIIRS) illustrated with data from landslide-prone Taiwan. We tested two hypotheses: 

\begin{enumerate}
\item Whether using multi-seasonal imagery yields a statistically more accurate classification of landslides compared to single-season imagery, and
\item Whether adding nighttime light data results in a more accurate classification of landslides. Further, we evaluated the potential transferability of the developed model, and retrospectively mapped landslides from the years 1998 to 2017 across Taiwan. 
\end{enumerate}

\section{Study area}

\begin{figure}
    \centering
        \includegraphics[width=0.95\columnwidth]{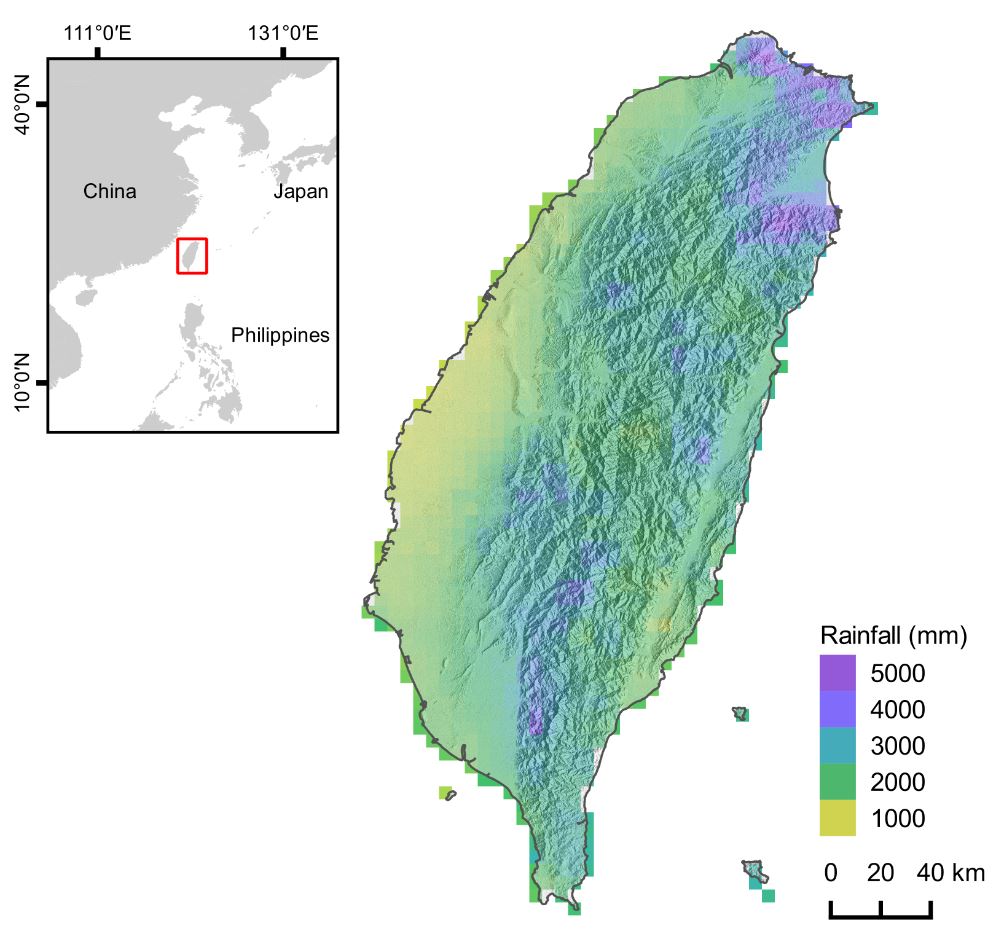}
    \caption{Map of Taiwan with hill shade and the average annual precipitation of 1963–2012.
(Source of rainfall data: TCCIP).}
    \label{fig:1}
\end{figure}

Our study area is mainland Taiwan with an area of 35,874 km\textsuperscript{2}, located in the Western Pacific Ring of Fire, with frequent earthquakes and comprising 75\% mountainous regions (Fig.~\ref{fig:1}). Despite a rugged terrain, it is densely populated with 610 people per km\textsuperscript{2}. Hilly terrain, earthquakes and precipitation are the main underlying drivers of landslides. Taiwan has an average annual precipitation of 2,500mm, varying between mountainous regions (max 4,800mm) and plains (min 1100mm) (Fig.~\ref{fig:1}). Due to its geographical positioning in a sub-tropical zone in the Pacific, Taiwan annually suffers from on-land tropical cyclone strikes (so-called typhoons) three-four times a year. Among the historical typhoons, disasters related to Typhoon Morakot (7 to 9 August 2009) produced an accumulated rainfall of 2,777 mm (\cite{Ge2010}), landslides and debris flows, resulting in nearly 700 deaths and approximately 4.7 billion USD in damage.

\section{Methods} 
\begin{figure*}[!t]
     \centering
     \includegraphics[width=0.7\textwidth]{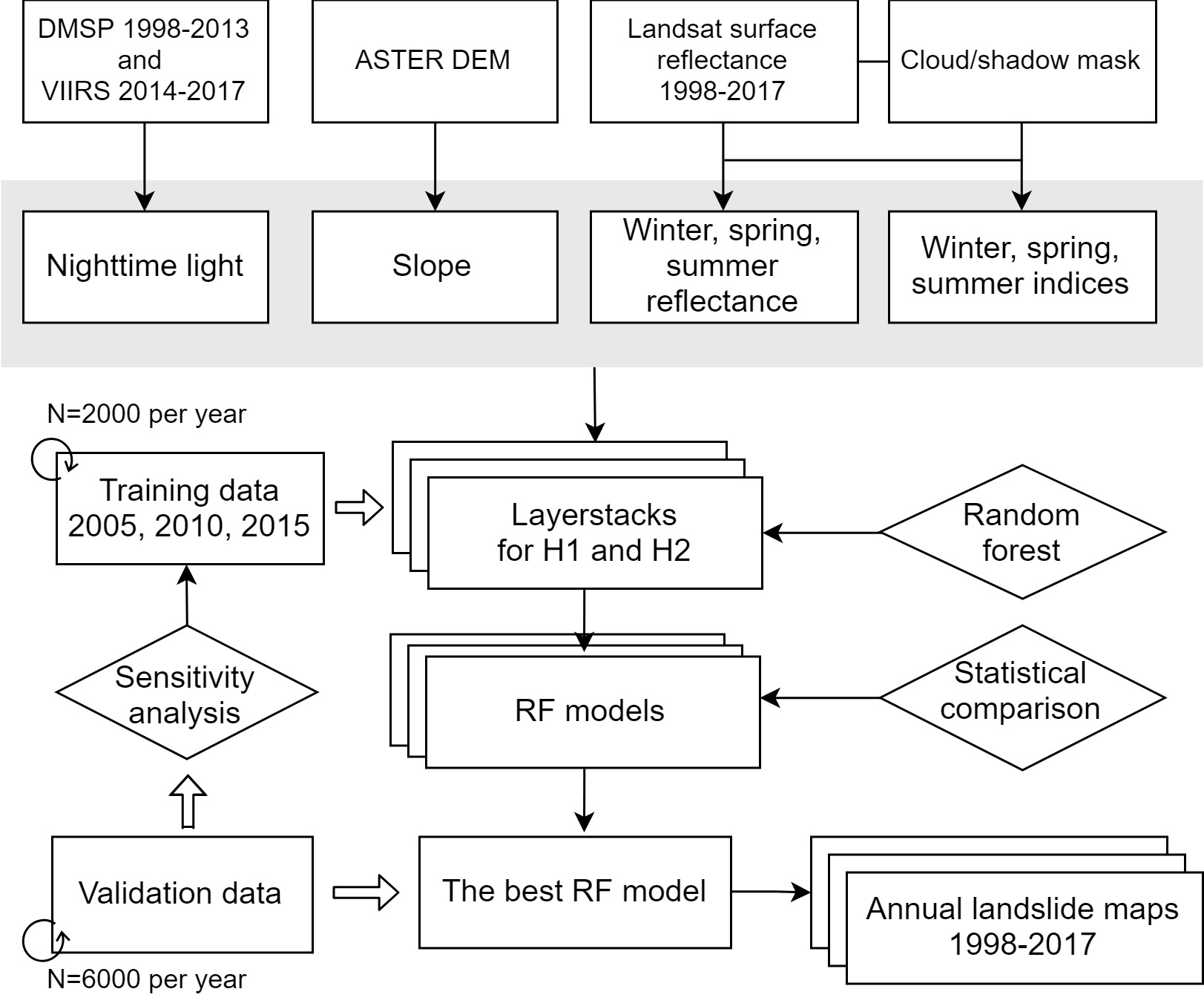}
     \caption{Flowchart of long-term landslide mapping based on open-access satellite datasets.}
     \label{fig:2}
\end{figure*}

The workflow entailed four steps to map landslide occurrence on an annual basis (Fig.~\ref{fig:2}). First, we extracted several remote sensing variables from Landsat time-series, DEM, and nighttime light data. We used cloud-free pixels and developed pixel-based composites of images (layerstacks) (Section~\ref{chap:Data}). To adopt a supervised classification, we created training and validation sample sets from visually validated reference maps (Section~\ref{chap:Training}). Second, to develop an accuracy-balanced random forest model, we conducted a sensitivity analysis of class ratios (the optimal ratio of training samples) (Section~\ref{chap:Sensitivity}). We used layerstacks with different sets of inputs (seasonal features, nighttime light, and slope), statistically tested our two hypotheses, and then selected the best model (Section~\ref{chap:Classification}). Thirdly, we validated with independent samples the applicability of the best model for other years with subsampled replicates (Section~\ref{chap:Accuracy}). Finally, we applied the RF model to map annual landslide occurrence explicitly from 1998 to 2017, demonstrating how these maps can be used to identify spatiotemporal patterns of landslide activities (Section~\ref{chap:Characterizing}).

\subsection{Datasets} \label{chap:Data}
Three datasets were utilized to derive variables for landslide classification (Table~\ref{tab:1}): (1) seasonal spectral features, such as NDVI, were derived from Landsat multispectral images, (2) slope was derived from the Global Digital Elevation Model of the Advanced Spaceborne Thermal Emission and Reflection Radiometer (ASTER GDEM) and (3) nighttime light data were based on DMSP and VIIRS. The variables derived from these datasets are listed in Table~\ref{tab:1}. All the variables were resampled to 30m pixel size with the nearest neighbor method.

\subsubsection{Landsat data} 
We downloaded from the USGS archive all available Level-2 Landsat-5, Landsat-7 and Landsat-8 imagery from 1998 to 2017 for the five Landsat tiles covering Taiwan (World Reference System WRS-2 path/row 117/43, 117/44, 117/45, 118/43, 118/44). Level-2 products are surface reflectance data at a 30m spatial resolution with systematic terrain correction and atmospheric correction, and are provided with per-pixel quality information, including a cloud mask produced with the CFMask algorithm (\cite{Foga2017}). We used the quality band to remove middle- and high-confidence cloud pixels. Areas of gaps caused by the scan line corrector failure (known as SLC-off) of Landsat-7 ETM+ were not processed for surface reflectance (\cite{Markham2004}). This step allowed us to avoid value anomalies but increased data loss, which was pronounced along the edge of the scenes, covering the middle part of our study area. We utilized the blue, green, red, near infrared, first and second shortwave infrared bands. NDVI and NDWI were first calculated for each single-date image (see formulas in Table~\ref{tab:1}). Then, the values of the bands and the indices were calculated as seasonal means in winter (December of the previous year to February), spring (March to May), and summer (June to August). This fusion step increased cloud-free observations (Fig. S1, Appendix A). We excluded images from September to November (Autumn), considering the fluctuation of typhoon perturbation and short-term vegetation recovery. 

\subsubsection{Aster GDEM data} 
The ASTER GDEM (obtained from \url{https://gdex.cr.usgs.gov/}) is a product of NASA and METI and was generated from ASTER imagery with a resolution of 30m. The elevation is assumed to be static in this study and we used the only single-time product of GDEM that is produced from stereo-pairs acquired from 2000 to 2011 to represent the general topography. Elevation information was used to derive the slope variable. 

\subsubsection{Nighttime light data} 
Nighttime light datasets include cloud-free DMSP data (1998-2013) and VIIRS data (2014-2017) and were downloaded from \url{https://www.ngdc.noaa.gov/eog/}. The DMSP annual composite data contain average radiance values of cloud-free coverages (Fig. S1, Appendix A), reflecting the persistent lights from cities, villages, and roads, with a spatial resolution of about 900m, and a temporal coverage of 1992 to 2013 (\cite{Elvidge2017,Imhoff1997}. Starting in 2013, VIIRS data is available with a finer spatial resolution of 450m approximately. As this newer data are only available in monthly composites (in contrast to the DMSP data which were aggregated to annual data), we used March data, with less cloud cover (Fig. S1, Appendix A) and the least number of Taiwanese holidays reducing in and out-migration, thus ensuring relatively stable lighting. To make VIIRS data comparable with DMSP data, we adopted a two-step procedure to spectrally stretch VIIRS data by the products in the overlapping year of 2013 for the study area. First, we removed outliers by forcing values greater than 40 and lower than 0.2 to 40 and 0.2, respectively. Second, we correlated the two datasets using a logarithmic transformation, producing a correlation coefficient of 0.94. We thus used a linear-log regression to stretch outlier-removed VIIRS data from 2014 to 2017 as follows:

\vspace{5mm}
$V' = 26.139 \times \log_{10}(V) + 23.179$
\vspace{5mm}

where $V'$ is the stretched VIIRS data, $26.139$ is the slope, $V$ is the outlier-removed VIIRS data, and $23.179$ is the intercept.

\subsection{Training and validation data collection } \label{chap:Training}
To produce training and validation datasets, we used reference landslide maps (2005-2016) from the Taiwan Council of Agriculture (data available from \url{https://www.tgos.tw}, also presented in Fig. S3, Appendix A). The twelve annual landslide maps covering the whole area of Taiwan were created with semi-automatic image classification approach (multispectral Formosat-2 satellite images at 2m resolution were taken between January and July). Non-vegetated areas that are not landslides were manually corrected, including human land use, river sediments, and displaced materials (debris/rock) overlying along gentle slopes. The minimum size of detected landslide objects was 1,000 m\textsuperscript{2}. These reference maps were validated by visual interpretation of national aerial photos with 98\% of overall accuracy (\cite{Lin2013formosat}). 

Training and validation sample sets were produced for ‘landslide' and ‘non-landslide' classes. We created sample points with an account for spatial autocorrelation, as spatial autocorrelation of the sample may reduce the representativeness of sampling (\cite{Millard2015}). We measured spatial autocorrelation with Moran's I (\cite{Ord2004}) and found that the spatial autocorrelation rapidly decreased (from 0.9 to 0.7) between 0-3 pixels. Thus, we sampled points by keeping the minimum distance of 100m ($>$ 3 pixels). Using this distance limit, the first sets of training and validation data were produced independently without overlap (validation points at the distance of 0-100 m from training sample were removed). The training set included 60,000 points for each of training years (2005, 2010, and 2015), and validation set included around 110,000 points for each of test years (2005-2016) (Table~\ref{tab:2}). The two sets were used for sub-sampling (drawing a small number of points) when classification and accuracy tests were iteratively implemented.

\vspace{0.1cm}
\begin{table}
\caption{Size of training and validation sets comprising points from different years. Points with missing values were removed. The two sample sets were used for iterative cross-validation.} 
\centering
\setlength{\tabcolsep}{5pt}
\centering
\label{tab:2}
\resizebox{\columnwidth}{!}{%
\footnotesize
\begin{tabular}{lllll}
\hline
\multirow{2}{*}{Year} & \multicolumn{2}{l}{Training   set} & \multicolumn{2}{l}{Validation   set} \\ \cline{2-5} 
                      & Landslide      & Non-landslide     & Landslide       & Non-landslide      \\ \hline
2005                  & 5961           & 49830             & 3901            & 105415             \\
2006                  & -              & -                 & 4413            & 108185             \\
2007                  & -              & -                 & 4409            & 112879             \\
2006                  & -              & -                 & 4663            & 107630             \\
2009                  & -              & -                 & 2836            & 111631             \\
2010                  & 13309          & 49758             & 8124            & 105846             \\
2011                  & -              & -                 & 6396            & 103638             \\
2012                  & -              & -                 & 6587            & 103198             \\
2013                  & -              & -                 & 6195            & 108040             \\
2014                  & -              & -                 & 6535            & 106919             \\
2015                  & 8676           & 49657             & 6505            & 111444             \\
2016                  & -              & -                 & 4966            & 111576             \\ \hline
\end{tabular}}
\end{table}

\subsection{Sensitivity analysis} \label{chap:Sensitivity}
We noticed that the class distribution was unbalanced, thus potentially introducing a bias of classification with overestimation of the dominant (non-landslide) and underestimation of the rare class (landslide) (\cite{Dalponte2013}). To evaluate the impact of unbalanced class distribution on the accuracy, we applied a sensitivity analysis approach proposed by \cite{Stumpf2011} that tests different class ratios ($\beta$) to define the optimal ratio of training samples to find an optimal balance between user's (overestimation) accuracy and producer's (underestimation) accuracy. For the analysis of class ratio optimization and the final classification and accuracy assessment, a 50-fold-iterative procedure was performed for each model. For each iteration, we adjusted the Stumpf and Kerle approach, by using a consistent total sample size to avoid a confounding sample size effect. In this way, 2000 training points for 2005, 2010, and 2015 were randomly drawn from the large training set, and 6000 validation points for each of the test years from the validation set were subsampled as test set (Fig.~\ref{fig:2}). The parameter $\beta_i$ refers to the ratio of landslide and non-landslide samples in the training set and was changed iteratively to approach an optimal value $\beta_n$ balancing producer's and user's accuracies. The procedure started from an equal class distribution ($\beta_i=1$), which defines the sample size for landslide ($2000{}/(1+\beta_i)$) and non-landslide ($2000\times \beta_i{}/(1+\beta_i)$), and in each step $\beta_i$ was increased by 0.1 until reaching a non-landslide-fivefold class distribution ($\beta_i=5$). To help the classifier to better learn the complexity of non-landslide activities, we ensured a minimum number of points (N=100) for six non-landslide land covers (building, road, agriculture, forest, non-landslide barren land, and water). In each iteration, 100 points were first allocated to every land cover class and then the remaining points were proportionally allocated.

\vspace{0.1cm}
\begin{table*}[!b]
\caption{Variables used for hypothesis 1 and 2.} 
\setlength{\tabcolsep}{5pt}
\centering
\label{tab:3}
\resizebox{\textwidth}{!}{%
\begin{tabular}{lll}
\hline
Classification scenario          & Target layerstack                                                   & Compared layerstack                                \\ \hline
(H1) Multiple seasonal imageries & Bands, indices in winter, spring, summer and slope                  & Bands, indices in a single season and slope        \\
(H2) Nighttime light data        & Bands, indices in winter, spring, summer, slope and nighttime light & Bands, indices in winter, spring, summer and slope \\ \hline
\end{tabular}}
\end{table*}

\subsection{Classification methods} \label{chap:Classification}

To map landslides, we used Random Forest (RF), a non-parametric machine learning classifier that has proven to accurately differentiate spectrally complex classes (\cite{Belgiu2016}). A RF classifier is an ensemble classifier that grows multiple decision trees and are trained using bagging, thereby letting the trees determine the probability of the class membership. RF performs multiple criteria classification, being at the same time fast and insensitive to overfitting (\cite{Breiman2001}). We utilized a version of RF implemented in the R statistical software with the randomForest package (\cite{Liaw2002}). 

RFs were trained with a combined training data subset from 2005, 2010, and 2015, which was class-ratio-adjusted. For parameterizing the RF classifier, based on out-of-bag accuracy, we used 500 trees and took the square root of the number of layers as a split criterion at each node. We ran two classification scenarios (Table~\ref{tab:3}) to test the two hypotheses that multi-seasonal imagery (H1) and nighttime light (H2) results in higher classification accuracy. To test H1, we classified a layerstack including bands and indices from three seasons (winter, spring and summer) and compared with the classification outcomes when only one single season was used. To test H2, we classified a layerstack with and without the additional inclusion of nighttime light data. We first separately estimated the optimal class ratio ($\beta$) for the H1 and H2 models by 50 iterations for each model with randomly subsampled training and validation datasets (with replacement). Based on their optimal numbers, we assessed the accuracies and statistically compared producer's and user's accuracies of the models for H1 and H2 using t-test at alpha=0.001.

\subsection{Accuracy assessment} \label{chap:Accuracy} 

To test whether the final model trained with the $\beta_n$-adjusted sample from the three years (2005, 2010 and 2015) is robust and applicable to other years, we used validation samples from twelve years (2005-2016) to evaluate annual accuracies. Although a random forest classifier calculates out-of-bag cross-validation accuracy, we relied on independent validation datasets for the final accuracy assessment. We subsampled 6,000 points for each of the tested years. Since the equal allocation of validation samples for unbalanced classes is not appropriate for area estimation and proportional allocation may result in imprecise estimates of user's accuracy for the rare class (a landslide in our case), we adopted an alternative sample allocation approach proposed by \cite{Olofsson2014}. We examined the standard errors of the estimated user's accuracy and the estimated area and allocated 500 points for a landslide, and the remaining 5500 points were allocated to the non-landslide class. 

These points were evaluated as one of four validation categories: true positive (TP), true negative (TN), false negative (FN), false positive (FP). TPs represent spatially and temporally correct landslide points, TNs are correct non-landslide points, FNs are reference landslide points missing in the detection and FPs are points identified as landslide but are not present in the reference map. Based on these categories, we calculated overall accuracy ((TP + TN / (TP + TN + FP + FN)), user's accuracy (TP / (TP + FP)), and producer's accuracy (TP / (TP + FN)). The user's accuracy represents the model performance in reducing landslide overestimation and producer's accuracy refers to the ability to minimize landslide underestimation. We implemented 100 subsampled replications to estimate the mean and standard deviation for the overall accuracy, user's accuracy and producer's accuracy for each test year. 

\begin{figure*}[!t]
     \centering
     \includegraphics[width=0.7\textwidth]{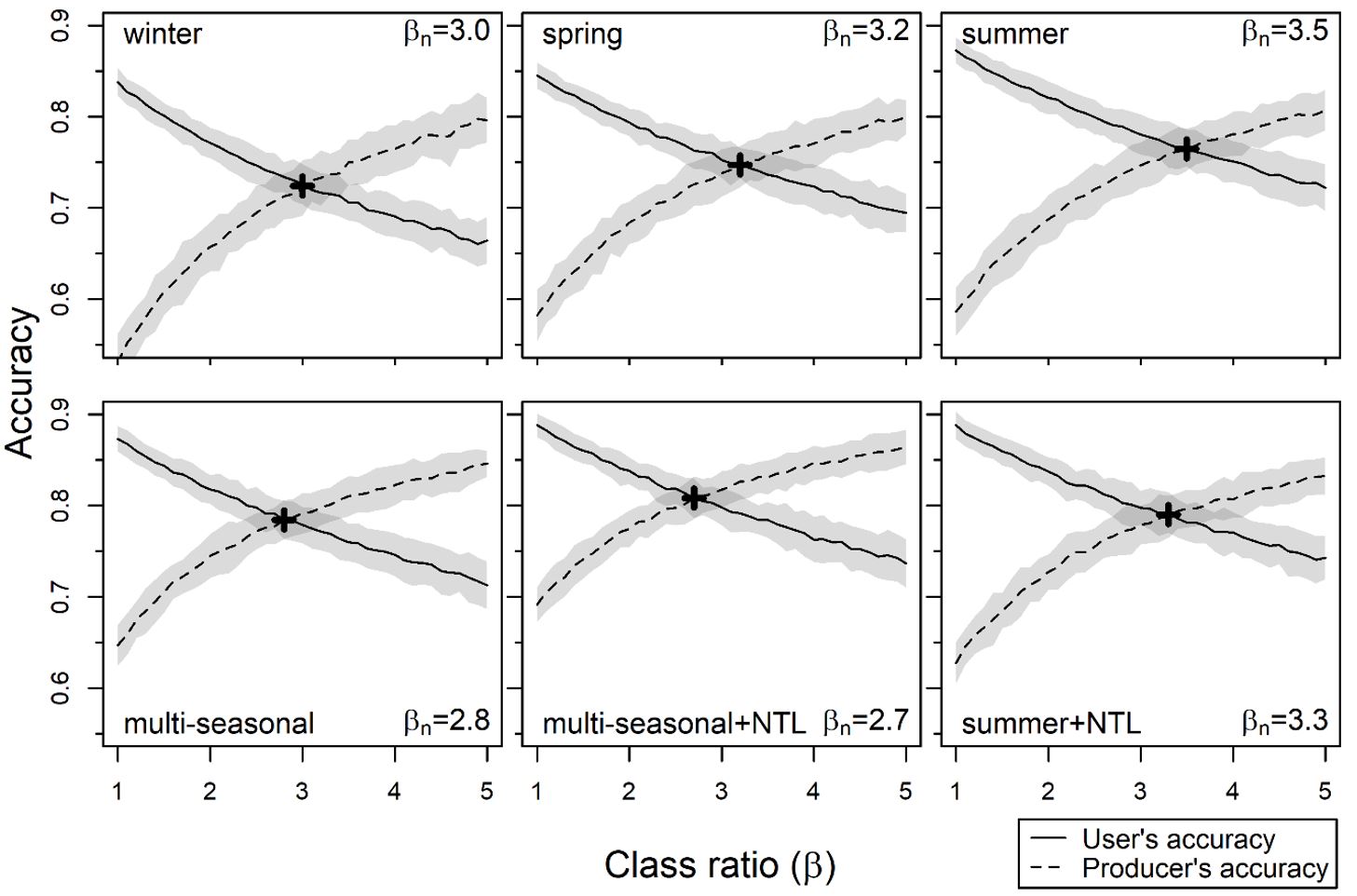}
     \caption{Estimates of the optimal class ratio ($\beta_n$) that achieves a balance between user's accuracy (solid line) and producer's accuracy (dot line). The lines indicating the
mean of the accuracy were produced from 50-fold subsampling runs for each $\beta_i$. The grey margins show 95\% confidence intervals.}
     \label{fig:3}
\end{figure*}

\subsection{Deriving spatiotemporal patterns of landslide activities} 
\label{chap:Characterizing} 
Based on the best RF model (Section~\ref{chap:Classification}) we derived the annual landslide / non-landslide maps from 1998 to 2017. To characterize the different types of long-term risk of landslides, we calculated the frequency, first occurrence, persistence and reoccurrence landslide metrics from the derived maps. First occurrence is the year the landslide was first observed from 1998 to 2017. Persistence expresses the longest number of consecutive years of landslide occurrence at a location. The frequency metric is a landslide rate (occurrences from 1998 to 2017 divided by the number of years). The reoccurrence metric addresses the number of landslide intervals. An interval ended if a pixel experienced revegetation (becoming non-landslide) in a year after landslide occurrence.

\section{Results}
We present our results in four parts: the sensitivity analysis (Section~\ref{res:Sensitivity}) and model comparison (Section~\ref{res:Effects}), which suggested that multi-seasonal imagery and nighttime light resulted in statistically significantly more accurate classification accuracies (p$<$0.001). When those features were used, the validation of the final model showed that user's and producer's accuracies over test years ranged from 73\% to 86\% (Section~\ref{res:Transferability}). In last Section~\ref{res:Characterizing}, the application results of detecting spatiotemporal patterns with the developed model were obtained.

\begin{figure*}[!b]
    \centering
        \includegraphics[width=0.7\textwidth]{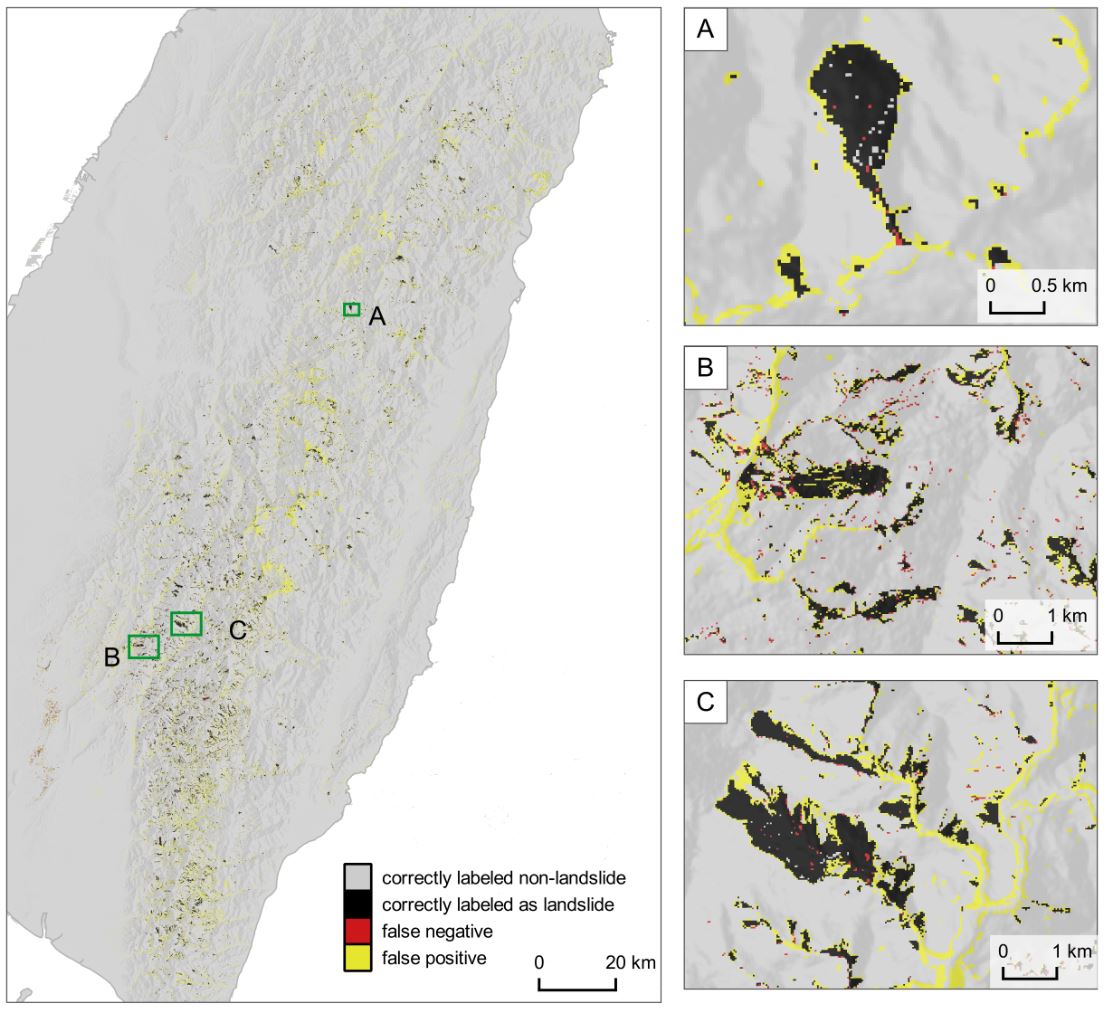}
    \caption{Final map based on the best model (classified multi-seasonal imagery combined with nighttime light data) for the year 2011, in comparison with a reference map. A: landslides on north-south slopes. B: a deep-seated landslide occurred at a village during the typhoon Morakot. C: large-scale landslides. False positives (yellow) from the displaced rock and debris retained as they were classified without manual elimination.}
    \label{fig:4}
\end{figure*}

\subsection{Sensitivity analysis}
\label{res:Sensitivity}
For all six models, we observed a strong over-estimation of landslide areas if samples of the two classes were equally sized ($\beta=1$) (Fig.~\ref{fig:3}). The over-estimation of landslide areas was even more pronounced if single-season models were utilized. By iteratively increasing $\beta_i$ the balance of user's and producer's accuracies was found to vary from 2.7 to 3.5 across the six models (Fig.~\ref{fig:3}). Since the optimal class ratio was model-dependent, we used the optimal ratio for each model independently for further model comparisons.

\vspace{0.1cm}
\begin{table*}[!b]
\caption{Statistical comparison of the models to test Hypothesis 1 (multi-seasonal imagery versus single-season imagery). Mean of accuracies for each case was estimated by runs of subsampling and based on identification of the optimal class ratio ($\beta$). T-tests were run to test whether the mean of accuracies for each model was statistically more accurate compared to the base model (winter).}
\centering
\setlength{\tabcolsep}{15pt}
\centering
\label{tab:4}
\resizebox{1\textwidth}{!}{%
\begin{tabular}{llllllllll}
\hline
\multirow{2}{*}{Model} & \multirow{2}{*}{$\beta_n$} & \multirow{2}{*}{Overall accuracy} & \multicolumn{3}{l}{User’s}            & \multicolumn{4}{l}{Producer’s}                            \\ \cline{4-10} 
                       &                            &                                   & accuracy & t value & p-value          & \multicolumn{2}{l}{accuracy} & t value & p-value          \\ \hline
Winter                 & 3.0                        & 95.4\%                            & 72.2\%   & —       & —                & \multicolumn{2}{l}{72.6\%}   & —       & —                \\
Spring                 & 3.2                        & 95.8\%                            & 74.5\%   & 8.5     & \textless{}0.001 & \multicolumn{2}{l}{74.9\%}   & 11.6    & \textless{}0.001 \\
Summer                 & 3.5                        & 96.1\%                            & 76.4\%   & 15.1    & \textless{}0.001 & \multicolumn{2}{l}{76.6\%}   & 18.3    & \textless{}0.001 \\
Multiple seasons       & 2.8                        & 96.4\%                            & 78.5\%   & 23.2    & \textless{}0.001 & \multicolumn{2}{l}{78.4\%}   & 26.1    & \textless{}0.001 \\ \hline
\end{tabular}}
\end{table*}

\vspace{0.1cm}
\begin{table*}[!b]
\caption{Statistical comparison of the models to test Hypothesis 2. T-tests were run to test whether the mean of accuracies of multi-seasonal imagery combined with nighttime light data was statistically more accurate compared to classified multi-seasonal imagery alone.}
\centering
\setlength{\tabcolsep}{15pt}
\centering
\label{tab:5}
\resizebox{1\textwidth}{!}{%
\begin{tabular}{lllllllll}
\hline
\multirow{2}{*}{Model}                                                            & \multirow{2}{*}{$\beta_n$} & \multirow{2}{*}{Overall accuracy} & \multicolumn{3}{l}{User’s}            & \multicolumn{3}{l}{Producer’s}        \\ \cline{4-9} 
                                                                                  &                            &                                   & accuracy & t value & p-value          & accuracy & t value & p-value          \\ \hline
\begin{tabular}[c]{@{}l@{}}Non-nighttime light \\ (multiple seasons)\end{tabular} & 2.8                        & 96.4\%                            & 78.5\%   & —       & —                & 78.4\%   & —       & —                \\
\begin{tabular}[c]{@{}l@{}}Nighttime light \\ (multiple seasons)\end{tabular}     & 2.7                        & 96.8\%                            & 80.8\%   & 10.5    & \textless{}0.001 & 80.8\%   & 10.8    & \textless{}0.001 \\ \hline
\end{tabular}}
\end{table*}

\subsection{Effects of using nighttime lights and multi-seasonal imagery}
\label{res:Effects}
Results of the t-test for scenario (H1) (Table~\ref{tab:4}) show that the multi-season model that includes the variables from all three seasons achieved balanced user's and producer's accuracies of 78\%, and performed statistically significantly more accurate (p$<$0.001) compared to single-season models for winter, spring and summer. Among the single-season models, summer and spring yielded statistically more accurate classification accuracies (user's and producer's accuracies range of 74-76\%) than the winter model, with user's and producer's accuracy around 72\%. For scenario (H2) (Table~\ref{tab:5}), combined nighttime light data and multi-seasonal satellite imagery resulted in a statistically more accurate classification accuracy (balanced user's accuracy and producer's accuracy of 81\%), compared to classified multi-seasonal satellite imagery alone (of 78\%). In the best model, nighttime light and slope have higher importance than other variables (Fig. S2, Appendix A).

\begin{figure}[!t]
    \centering
    \resizebox{1\columnwidth}{!}{\includegraphics{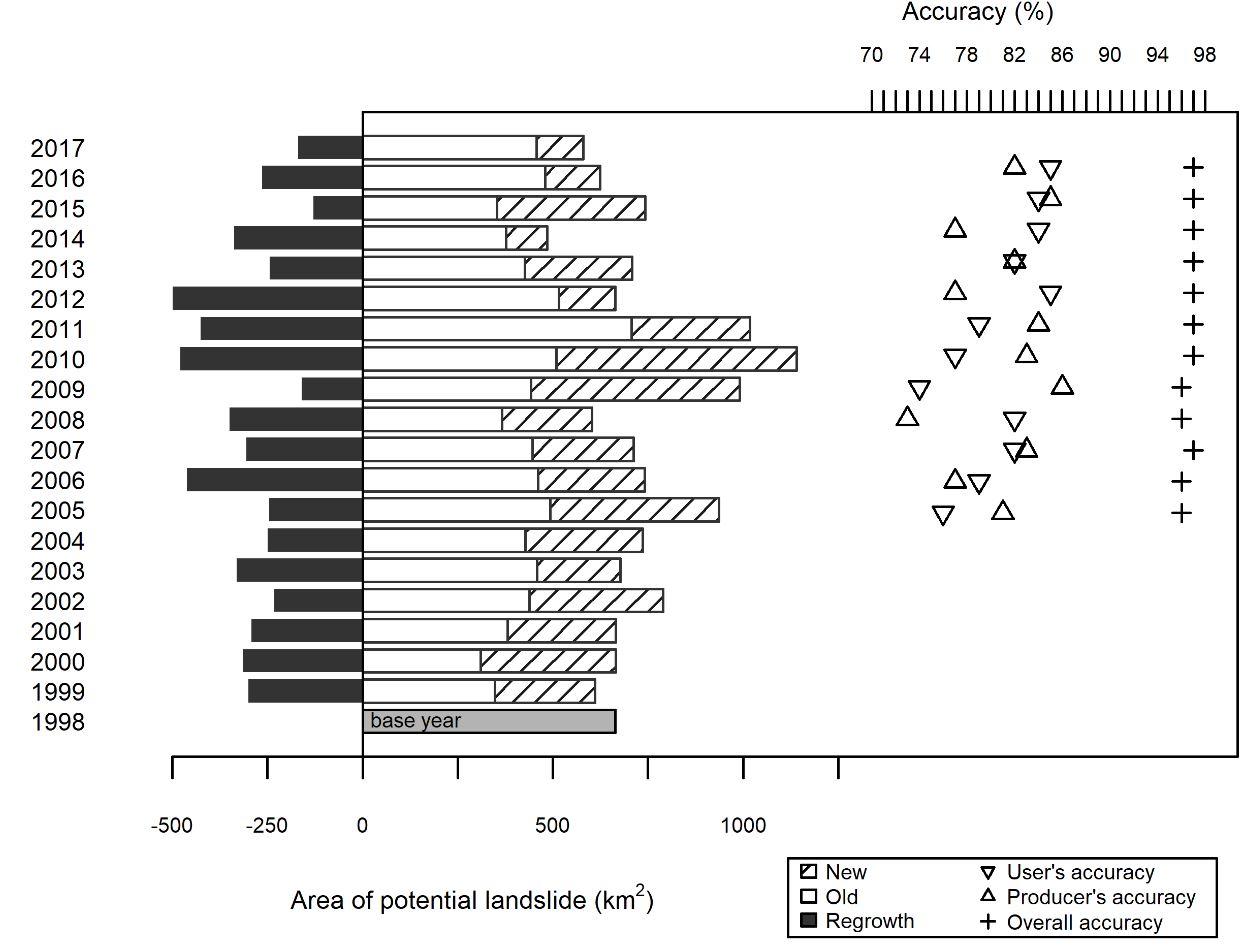}}
    \caption{Estimated landslide area split into old landslide area (year t-1 and year t: landslide), new landslide area (year t-1: non-landslide and year t: landslide), and
revegetated area (year t-1: landslide and year t: non-landslide), with user's accuracies, producer's accuracies, and overall accuracies from validated maps 2005–2016.
The area (km2) is adjusted for missing areas from clouds and SLC-off gaps.}
    \label{fig:5}
\end{figure}

\subsection{Transferability of the model over time}
\label{res:Transferability}

When applying the best model, including variables of nighttime light, slope, and multiple seasonal bands, NDVI, and NDWI, our accuracy assessment results showed that the overall accuracies in other years ranged from 96\% to 97\% (Table~\ref{tab:6}). The user's accuracies varied between 74 and 85\% and producer's accuracies were between 73 and 86\%, implying that the model can be applied also for other years. We observed that user's accuracy in years 2014 to 2016 were higher (average of 84\%) than previous years (average of 80\%) (Table 6), indicating that the higher resolution of nighttime light data helps to reduce over-prediction.

\begin{table}
\caption{Annual overall, user’s and producer’s accuracies (\%) based on different nighttime light (NTL) resolution. The mean values and 95\% confidence intervals (±) were estimated from 100 iterations with subsampling of 6000 samples.} 
\centering
\setlength{\tabcolsep}{10pt}
\centering
\label{tab:6}
\resizebox{\columnwidth}{!}{%
\begin{tabular}{lllll}
\hline
Year & Overall accuracy & UA & PA & NTL resolution \\ \hline
2005 & 96±1             & 76±4            & 81±4                &                \\
2006 & 96±1             & 79±3            & 77±4                &                \\
2007 & 97±1             & 82±3            & 83±4                &                \\
2008 & 96±1             & 82±3            & 73±4                &                \\
2009 & 96±1             & 74±3            & 86±3                & 900m           \\
2010 & 97±1             & 77±3            & 83±4                &                \\
2011 & 97±1             & 79±3            & 84±4                &                \\
2012 & 97±1             & 85±4            & 77±4                &                \\
2013 & 97±1             & 82±3            & 82±4                &                \\
2014 & 97±1             & 84±3            & 77±4                &                \\
2015 & 97±1             & 84±3            & 85±3                & 450m           \\
2016 & 97±1             & 85±3            & 82±4                &                \\ \hline
\end{tabular}}
\end{table}

The validated landslide map for 2011 showed the landslides at different locations, their sizes, and shapes (Fig.~\ref{fig:4}). Landslides dominated in the middle and southern parts of the central mountain range. A subset (Fig.~\ref{fig:4}A) illustrates how landslides were better depicted when landslide geometry was more compact. Although the compact and massive landslides were successfully detected, false positives were predominantly distributed along the edge of the small landslides, where mixed pixels are common. Inset (Fig.~\ref{fig:4}B) highlights the classified occurrence of a disastrous deep-seated landslide induced by extreme rainfall on 8 August 2010 – the “Shiaolin landslide” that destroyed a village, causing 491 fatalities. Deep-seated landslides are usually larger than shallow landslides. Fig.~\ref{fig:4}C presents a large-scale landslide at the vicinity of Fig.~\ref{fig:4}B. 

We identified the composition of existing landslides, which we classified into old ones, new ones, and also revegetated areas (Fig.~\ref{fig:5}). After the dramatic Chi-Chi earthquake in late 1999, the overall landslide area in 2000 was only 6\% higher compared to that prior to the earthquake. However, the estimate of new landslide areas indicated a significant increase by 34\% compared to the previous year, which shows how extreme this event was.

\begin{figure*}[!t]
    \centering
        \includegraphics[width=0.9\textwidth]{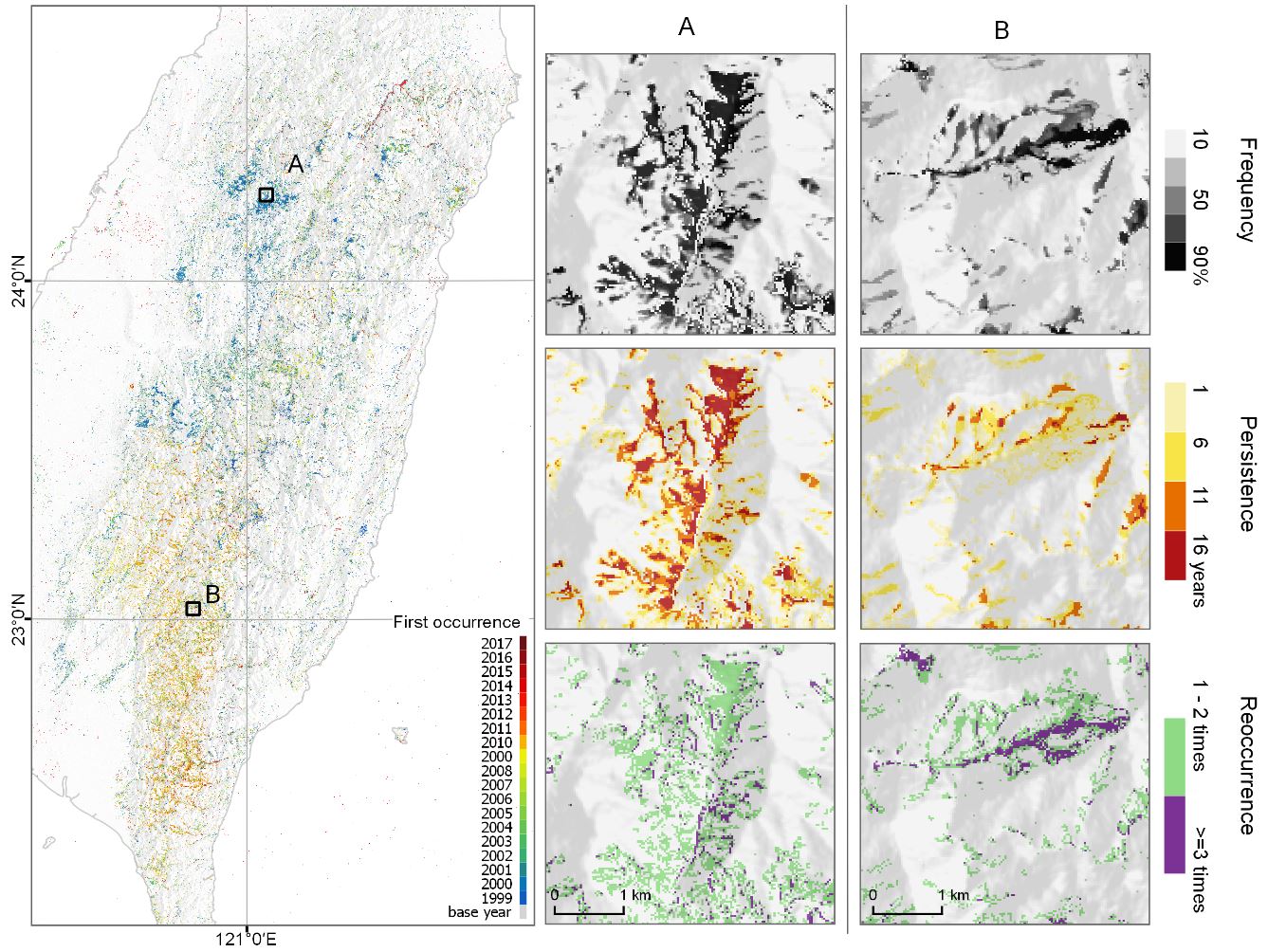}
    \caption{First occurrence of landslides from 1998 to 2017. (A) An earlier occurring area and (B) a post-Morakot landslide area. Both sites present high frequency, while (A) is of high persistence, and (B) is of high reoccurrence.}
    \label{fig:6}
\end{figure*}

\subsection{Characterizing spatiotemporal patterns of landslide activities}
\label{res:Characterizing}
We applied the most accurate model (incorporating variables of nighttime light, slope, and multiple seasonal bands, NDVI, and NDWI) for annual mapping landslides from 1998 to 2017 across Taiwan. The results showed that landslide areas varied from 1998 to 2017 across Taiwan (Fig.~\ref{fig:5}) with an increase in landslide occurrence in the south of Taiwan (Fig.~\ref{fig:6}). 

The landslides in the northern regions were mostly triggered before 2002 and after the Chi-Chi earthquake in 1999 (Fig.~\ref{fig:6}), while landslides in the southern regions predominantly occurred from 2009, corresponding to the timing of typhoon Morakot. The temporal patterns of landslide activities are different between a northern site (Fig.~\ref{fig:6}A) and a southern site (Fig.~\ref{fig:6}B), in terms of frequency, persistence, and reoccurrence.

\section{Discussion}
We developed an approach using freely available Landsat multi-seasonal imagery and nighttime light data to reconstruct long-term dynamics of landslides in Taiwan. Event-based commercial very high-resolution landslide maps are usually used for depicting morphological changes in size, length, and symmetry of the individual landslides. Yet, freely available medium resolution Landsat imagery and nighttime light data allows the long-term and large-scale assessment of landslide patterns, thus providing knowledge about frequency, persistence, and reoccurrence of landslide activities. Our study underscored the utility of the synergetic use of nighttime light data and multi-seasonal imagery to map landslides annually. With the metrics including multiple seasonal bands, NDVI, NDWI, slope, and nighttime light, the RF classifier reached an overall accuracy of 97\%, and a balanced user's and producer's accuracies of 81\%. When applying the developed RF model for other years without training samples for the respective years, we also achieved high overall, user's, and producer's accuracies (96-97\%, 74-85\%, and 73-86\%). The accuracies achieved are more balanced compared to other recent studies of large-scale landslide detection in Asia, with user's and producer's accuracy at 58\% and 87\% based on a multi-sensor approach (resolution at 5m-30m) for long-term landslide mapping (\cite{Behling2016}), and at 32\% and 63\% based on single-temporal Landsat imagery (\cite{Yu2017}). Our approach is comparable in accuracy to a study on landslide detection based on object-based classification of optical imagery and DEM at higher resolution (2-10m) which also reaches balanced accuracies between 73\% and 87\% (\cite{Stumpf2011}). 

We found nighttime light imagery helpful for increasing classification accuracy of landslides in anthropogenic areas. Despite its relative coarse resolution, luminosity from built-up areas and road lights allows the models to account for the human activities not affected by landslides and thus to separate them from landslide affected areas. Earlier studies have found the usefulness of nighttime light imagery to assess the changes in population density, various economic activities and technological shifts (\cite{Bennett2017,Rybnikova2017}). To our knowledge, the current study is the first to test the suitability of employing nighttime light imagery to study landslides. Our results suggest that the use of nighttime light data helps in disentangling the complex interrelationship between spectral signatures of natural and human-managed land cover classes, and therefore is of usefulness in mapping impacts of hazards. 

We found that for highly populated Taiwan multi-seasonal imagery resulted in more accurate classification of landslides than single-season imagery. The use of machine-learning random forest is powerful in separating multi-dimensional seasonal signals. For instance, built-up areas have stable reflectance across winter, spring and summer, while landslides may have fluctuating reflectance values across the year due to the contribution from the varying vegetation signals associated with regrowth. Several studies have proven the utility of multi-seasonal imagery and multiple criteria classifiers to accurately classify various land-cover types and land use (\cite{Oetter2001,Guerschman2003,Bleyhl2017,Gao2015,Gradinaru2017}). Our study highlights the power of multi-seasonal imagery for accurately mapping landslides. 

Landslides are widespread throughout the world but tend to have been mapped in details and only at a very localized scale or for a short duration (\cite{VanWesten2006}). The current global database of landslides records only fatal landslides (\cite{Petley2012}), and is not updated with past landslides and whether areas of landslides were recovered or rehabilitated. Taiwan has a very dynamic landslide environment, driven by high precipitation events, frequent earthquakes, and fast vegetation regrowth (\cite{Chang2007}). We found an inter-annual fluctuation of the total potential landslide area between 478 to 1130 km\textsuperscript{2} in Taiwan without a consistent increase or decrease between 1998 and 2017. This absence of a trend is in contrast to the continuing upward trend of globally recorded landslides (\cite{Petley2012}). We observed that landslide areas could largely recover within a year (Fig.~\ref{fig:5}), and that landslide areas decreased by 35\% from 2011 to 2012 and by 32\% from 2013 to 2014. In contrast, the highest growth rate from 2008 to 2009 (65\%), is explained by the tropical cyclone Morakot that hit Taiwan in August 2009, confirming the general trend observed by \cite{Petley2010} that the major driver of landslides in East Asia is tropical cyclones.

There were some limitations in our study, such as a necessity to fuse multi-sensor imagery. Multi-sensor images may have different spatial, spectral, temporal and radiometric resolutions (\cite{Pohl1998}) and for instance an intercalibration of DMSP and VIIRS nighttime light data is required. To mitigate the impacts of application of inconsistent data, we trained our model by selecting training years covering different sensors' span. Additionally, nighttime light data have a relatively coarse resolution (at best 500m) which differed from the 30-m Landsat imagery we utilized in our study. Nevertheless, a combination of multi-sensor data allows increasing the time span of observations, and facilitates highlighting specific phenomena, which may not be possible to detect with a single remote sensing product (e.g., optical imagery). 

Our landslide maps did not fully cover our study area, because of data availability of optical imagery in this cloud-prone region. Therefore, we missed detection of landslides in Taiwan for 13\% and 11\% of the total area in 1998 and 2013, respectively, as there were less accessible images (before Landsat-7 launched in 1999 and right after Landsat-5 stopped collecting data in 2013). Where image availability and cloud-contamination is an issue, the number of observations can be complemented with additional use of radar (e.g., Sentinel-1 since 2013) imagery, as well as fusion using images from MODIS and Landsat (\cite{Zhu2016}). We did not test the performance of higher resolution remotely sensed products that are also freely available, such as from the Sentinel-2 MSI time-series imagery. Sentinel-2 MSI imagery has high potential to enhance classification of landslides with higher spatial accuracy, but limited historical coverage. What our study showed is that, even with radiometrically coarse DMSP nighttime light data and optical Landsat imagery, it is possible to reconstruct historical landslides as early as the early 1990s accurately.

Despite some limitations, we propose the idea of using cost-effective data and methods to provide knowledge of historical landslide inventories, reflecting landslide first occurrence, persistence, as well as identify reoccurring landslide activities. Compared to areas showing only recent landslides, the areas of persistent landslides are more predictable and land-use plans around these regions can be formulated in advance. The reoccurrence rate is related to the depths of the sliding plane, which is useful for identifying remediation strategies. These annual landslide maps also provide a basis for studies on drivers of landscape changes and to help develop policy interventions.

\section{Conclusions}
Here we explored the effectiveness of nighttime light and multi-seasonal imagery to provide improved landslide image classification and for the derivation of long-term landslide maps in Taiwan from 1998 to 2017. Our findings confirmed that nighttime light and multi-seasonal optical imagery could significantly improve landslide mapping accuracies. This approach proved to be robust and transferable to historical mapping of landslides over a long period (20 years). We demonstrate the usefulness of long-term landslide maps that allow characterizing spatio-temporal patterns of landslide characteristics such as first time of occurrence, persistence and reoccurrence rate. Notably, during the period 1998 to 2017 the patterns of high landslide occurrence shifted from the north to the south of Taiwan. Our findings of the utility of nighttime light imagery and optical multi-seasonal imagery to map retrospective landslides can likely be replicated elsewhere, particularly because of the growing archives of freely available optical Landsat-like multi-seasonal imagery and refined VIIRS nighttime light data.

\section*{Acknowledgement}
This work was supported by a Ph.D. scholarship from the Ministry of Education, Taiwan, and by BERTHA - the Danish Big Data Centre for Environment and Health funded by the Novo Nordisk Foundation Challenge Programme (grant NNF17OC0027864). The authors wish to thank the three anonymous reviewers for very thorough and constructive reviews, and Tzu-Yin Kasha Chen for comments on geomorphological ex- planation.

\section*{References}
\printbibliography[title={references},heading=none]

\clearpage

\includepdf[pagecommand={\thispagestyle{plain}},pages=-, angle=0]{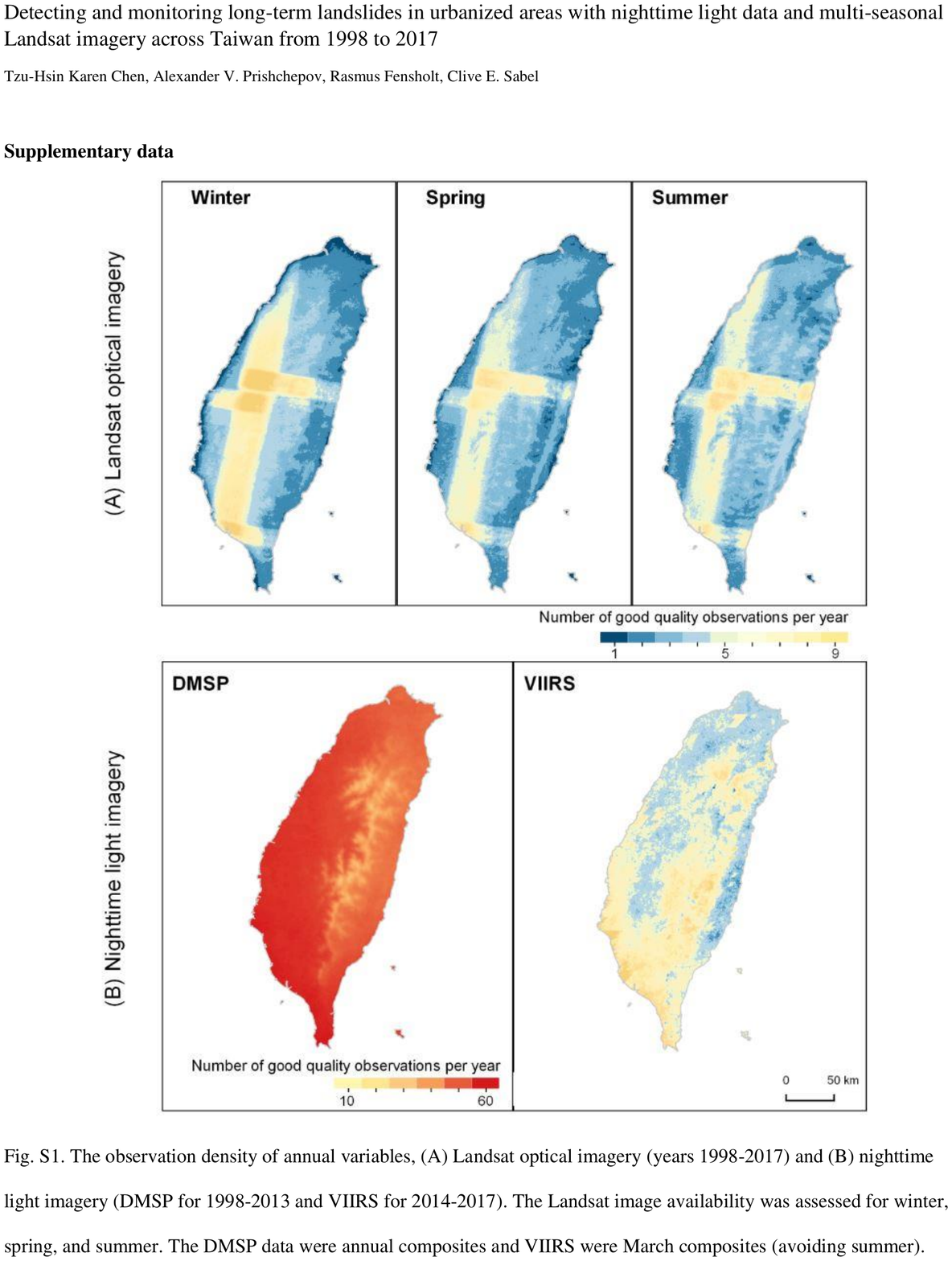} 

\end{document}